\documentclass{PoS}
\usepackage{amssymb,amsmath}

\title{Light-Meson Two-Photon Decays in Full QCD}
\ShortTitle{Light-Meson Two-Photon Decays in Full QCD}

\author{\speaker{Saul D. Cohen} \\
        Thomas Jefferson National Accelerator Facility, Newport News, VA 23606 \\
        E-mail: \email{sdcohen@jlab.org}}

\author{Huey-Wen Lin, Jozef Dudek, Robert G. Edwards \\
        Thomas Jefferson National Accelerator Facility, Newport News, VA 23606 \\}

\abstract{We present a study of two-photon decays of light mesons, focusing on the neutral pion decay. This important process highlights the effects of the axial anomaly in QCD but has been little studied on the lattice. By applying the Lehmann-Symanzik-Zimmermann (LSZ) reduction formula, we reconstruct the electromagnetic matrix elements from three-point vector-vector Green functions calculated on 2+1-flavor isotropic clover lattices.}
\FullConference{The XXVI International Symposium on Lattice Field Theory \\
		 2008 July 14--19 \\
		 Williamsburg, Virginia, USA}
		
\begin{document}

%%%%%%%%%%%%%%%%%%%%%%%%%%%%%%%%%%%%%%%%%%%%%%%%%%%%%%%%%%%%%%%%%%%%%%%%%%%%%%%%
\section{Introduction}
%%%%%%%%%%%%%%%%%%%%%%%%%%%%%%%%%%%%%%%%%%%%%%%%%%%%%%%%%%%%%%%%%%%%%%%%%%%%%%%%
The decay of the neutral pion into two photons has attracted a great deal of scrutiny since the early days of particle theory. In the Standard Model, this decay is understood to be primarily due to the effects of the Adler-Bell-Jakiw axial anomaly, where it is much larger than any competing process. Despite its long history, the $\pi^0\rightarrow\gamma\gamma$ decay is difficult to study experimentally, since both the initial and final states are electrically neutral. The large uncertainty in this width (7.84(56)~eV, about 7\%) propagates into determinations of the splitting between the up- and down-quark masses.

Given its importance as a fundamental prediction of the Standard Model and the relatively poor precision with which it is currently known, it is not too surprising that there is a substantial experimental program attempting to improve this measurement. The Primakoff Experiment (PrimEx) in Hall~B at Jefferson Lab (JLab) uses photon fusion between a real photon and the electric field of a heavy nucleus to produce neutral pions. Combined with a high-precision calorimeter, PrimEx was constructed to provide a percent-level precision measurement of the neutral pion width.
%\cite{PrimEx}.

From the theoretical side, the situation is somewhat better. The amplitude of the neutral pion to two photons process is exactly given by the axial anomaly in the chiral limit, and chiral perturbation theory (XPT) may be used to give a physical result. To leading order, XPT gives
\begin{equation}
\Gamma(\pi^0\rightarrow\gamma\gamma) = \frac{\alpha^2N_c^2M_\pi^3}{576\pi^3F_\pi^2} \approx 7.72\mbox{ eV},
\end{equation}
which is in apparent agreement with experiment. However, there are some complications. At next-to-leading order, XPT must consider the mixing of the pion with the eta and eta-prime. This increases the prediction to 8.10(8)~eV\cite{Goity:2002nn}.
Lattice QCD should also be able to determine this quantity from first principles, but since XPT appears to perform so well, this will simply be a crosscheck. On the other hand, lattice calculations can address other related quantities where XPT is expected to perform poorly, such as the processes $\eta\rightarrow\gamma\gamma$, $\eta^\prime\rightarrow\gamma\gamma$ and $a_0\rightarrow\gamma\gamma$. Measurements of these quantities will be performed by PrimEx following the 12~GeV upgrade at JLab. In addition, experiments using photoproduction (such as GlueX) will probe photon-fusion production of neutral mesons ($\gamma\gamma^*\to\Phi^0$).

Despite the clear motivations for making a measurement of two-photon decays of neutral mesons, lattice calculations have done little to address the issue. The difficulty faced in lattice QCD is simple: the photon is not an eigenstate of QCD. An operator carrying the quantum numbers of the photon in lattice QCD will instead create a rho meson (or two pions, depending on the quark mass used). The resolution of this difficulty, proposed by Ji and Jung\cite{Ji:2001wha,Ji:2001nf},
uses perturbative field theory techniques to express the photon as a superposition of QCD eigenstates accessible on the lattice. This technique was recently used to study the two-photon decays of charmonium in Ref.~\cite{Dudek:2006ut}.

\section{Lehmann-Symanzik-Zimmermann Reduction}

The method of Ji and Jung applies the Lehmann-Symanzik-Zimmermann (LSZ) reduction to evaluate a photonic matrix element. We are interested in matrix elements of the form
\begin{equation}
\langle \gamma(q_1,\lambda_1)\gamma(q_2,\lambda_2)|\Phi(p) \rangle,
\end{equation}
where the two photons have momenta $q_{1,2}$ and polarizations $\lambda_{1,2}$ and the neutral meson $\Phi$ has momentum $p$. Using a standard LSZ reduction, we re-express this using a time-ordered (for now, working in Minkowski spacetime) product of photon fields:
\begin{equation}
-\lim_{\substack{q_1^\prime \to q_1\\q_2^\prime \to q_2}} \epsilon^{(1)*}_\mu(q_1,\lambda_1) \epsilon^{(2)*}_\nu(q_2,\lambda_2) {q_1^\prime}^2 {q_2^\prime}^2 \int\!\!d^4x\,d^4y\,
  		e^{iq_1^\prime \cdot y + iq_2^\prime \cdot x}
  		\langle 0| T\{A^\mu(y) A^\nu(x)\} |\Phi(p) \rangle,
\end{equation}
where $\epsilon$ is a polarization tensor and $A$ is the vector photon field, which we have Fourier transformed.

Since lattice QCD can only directly measure eigenstates of QCD, we cannot treat these $A$ fields. One solution to this might be to include explicit QED U(1) fields during our lattice gauge generation, but since the cost of creating new ensembles is so great, we would rather attempt to express them in terms of quark and gluon operators accessible to QCD alone. To do this, we apply perturbative QED to integrate out the photon fields:
\begin{multline}
\int\!\!DA\,D\bar{\psi}\,D\psi\,e^{iS_{\rm QED}}A^\mu(y)A^\nu(x) \approx \\
\int\!\!DA\,D\bar{\psi}\,D\psi\,e^{iS_0}\left(...+\frac{e^2}{2}\left[\bar{\psi}\gamma^\rho\psi A_\rho\right](z)\left[\bar{\psi}\gamma^\sigma\psi A_\sigma\right](w)+...\right)A^\mu(y)A^\nu(x),
\end{multline}
where we take a path integral over all possible configurations of the photon field $A$ and the quark fields $\psi$ and $\bar{\psi}$ (with electric charge $e$) using the free-field action $S_0$ and the interacting QED action $S_{\rm QED}$. This can be simplified by applying Wick contraction to the photon fields:
\begin{multline}
-e^2 \lim_{\substack{q_1^\prime \to q_1\\q_2^\prime \to q_2}}
\epsilon_\mu^{(1)*} \epsilon_\nu^{(2)*}
			{q_1^\prime}^2 {q_2^\prime}^2 \int\!\!d^4x\,d^4w\,d^4z\,
  		e^{iq_1^\prime \cdot x} D^{\mu\rho}(0,z) D^{\nu\sigma}(x,w)
  		\langle 0| T\{j_\rho(z) j_\sigma(w)\} |\Phi(p) \rangle,
\end{multline}
where $j_\mu=\bar{\psi}\gamma_\mu\psi$ and $D$ is the photon propagator,
\begin{equation}
D^{\mu\nu}(0,z) = -ig^{\mu\nu} \int \frac{d^4q}{(2\pi)^4}\frac{e^{i q \cdot z}}{q^2 + i\epsilon}.
\end{equation}

By substituting the explicit form of the photon propagator, we cancel the inverse propagators and most of the integrals over spacetime become simple delta-functions. The remaining expression is
\begin{equation}
e^2 \epsilon_\mu^{(1)*} \epsilon_\nu^{(2)*} \int\!\!d^4y\, e^{iq_1 \cdot y}
  		\langle 0| T\{j^\mu(0) j^\nu(y)\} |\Phi(p) \rangle.
\end{equation}
We now need to rotate into Euclidean space. This is possible as long as our integration contour does not hit a pole, where one of the photons can mix with an on-shell particle. Therefore, we must keep $q^2 < M_\rho^2$ (or $E_{\pi\pi}^2$, depending on the quark mass).

The resulting expression is something that can be evaluated in lattice QCD:
\begin{multline}
\frac{e^2 \epsilon_\mu^{(1)} \epsilon_\nu^{(2)}}
         {\frac{Z_\Phi(p)}{2E_\Phi(p)}e^{-E_\Phi(p)(t_f-t)}}
         \int\!\!dt_i\,e^{-\omega_1(t_i-t)}
 \left\langle T\left\{\int\!\!d^3\vec{z}\,e^{-i\vec{p}\cdot\vec{z}}\varphi_\Phi(\vec{z},t_f)
              \int\!\!d^3\vec{y}\,e^{i\vec{q_2}\cdot\vec{y}}j^\nu(\vec{y},t)j^\mu(\vec{0},t_i)
  \right\} \right\rangle,
  \label{eq:3pt}
\end{multline}
where $\omega_1$ is the energy of the first photon, $\varphi_\Phi$ is a creation operator for the neutral meson and $Z_\Phi$ and $E_\Phi$ are its overlap factor and energy respectively. The expression between the angled brackets is just the three-point correlation function with a meson on one end and vector currents at the other end and inserted. The remaining parts describe how to combine these QCD currents into a photon of the appropriate energy.

%%%%%%%%%%%%%%%%%%%%%%%%%%%%%%%%%%%%%%%%%%%%%%%%%%%%%%%%%%%%%%%%%%%%%%%%%%%%%%%%
\section{Lattice Calculation}
%%%%%%%%%%%%%%%%%%%%%%%%%%%%%%%%%%%%%%%%%%%%%%%%%%%%%%%%%%%%%%%%%%%%%%%%%%%%%%%%

The most straightforward way to evaluate Eq.~\ref{eq:3pt} on the lattice is to compute the three-point function with all insertion times $t_i$ and perform the integral by explicit summation. For this first test of the light-meson two-photon decay calculation, we use lattices made available through the ILDG by CP-PACS. Particularly, we use $20^3\times40$ lattices with 2+1 dynamical flavors of clover fermions having pion mass $M_\pi=725$~MeV and lattice scale $a^{-1}\approx2.25$~GeV\cite{Aoki:2008sm}.

\begin{figure}[h!]
\includegraphics[width=0.475\textwidth]{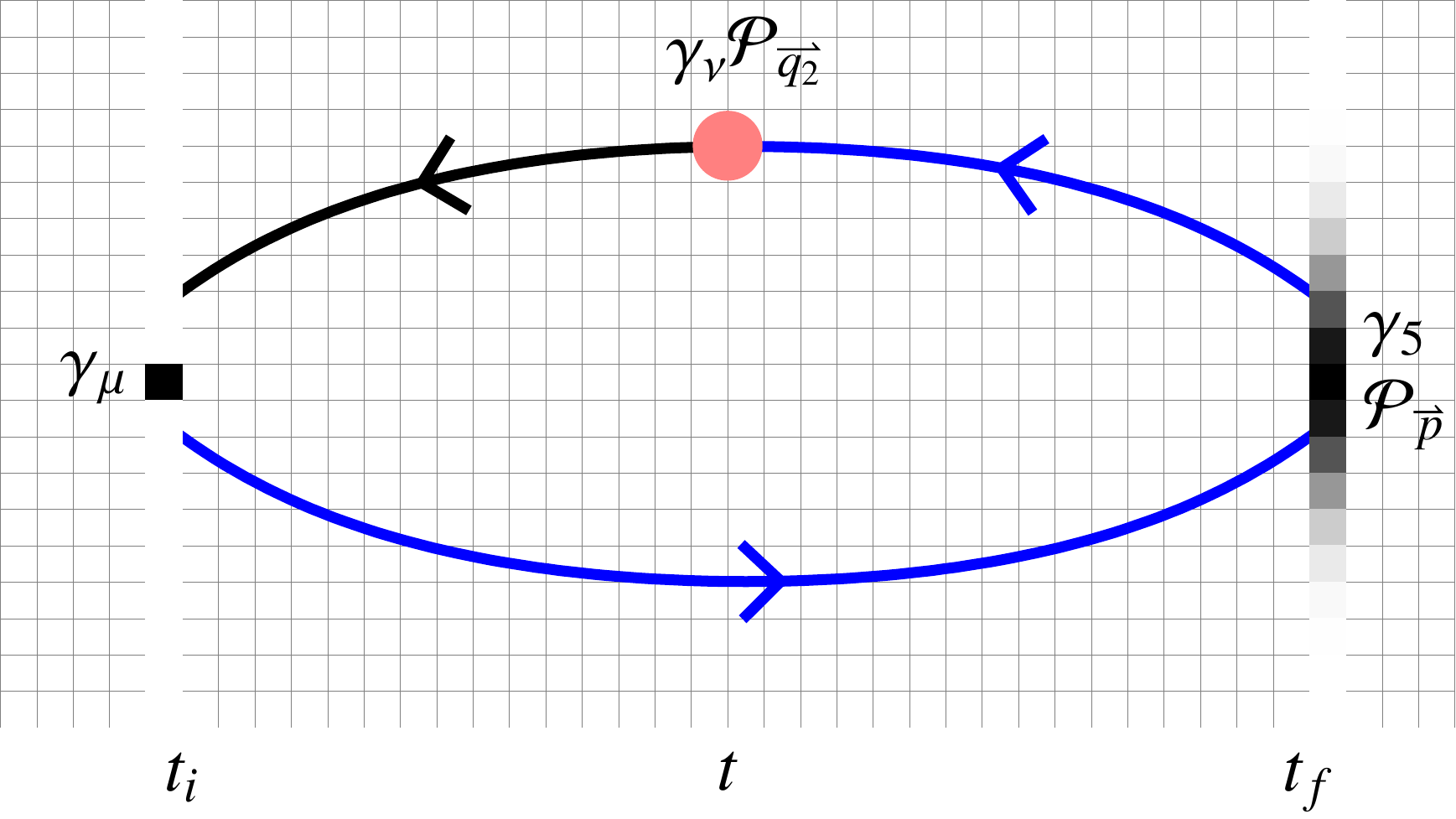}
\hspace{0.05\textwidth}
\includegraphics[width=0.475\textwidth]{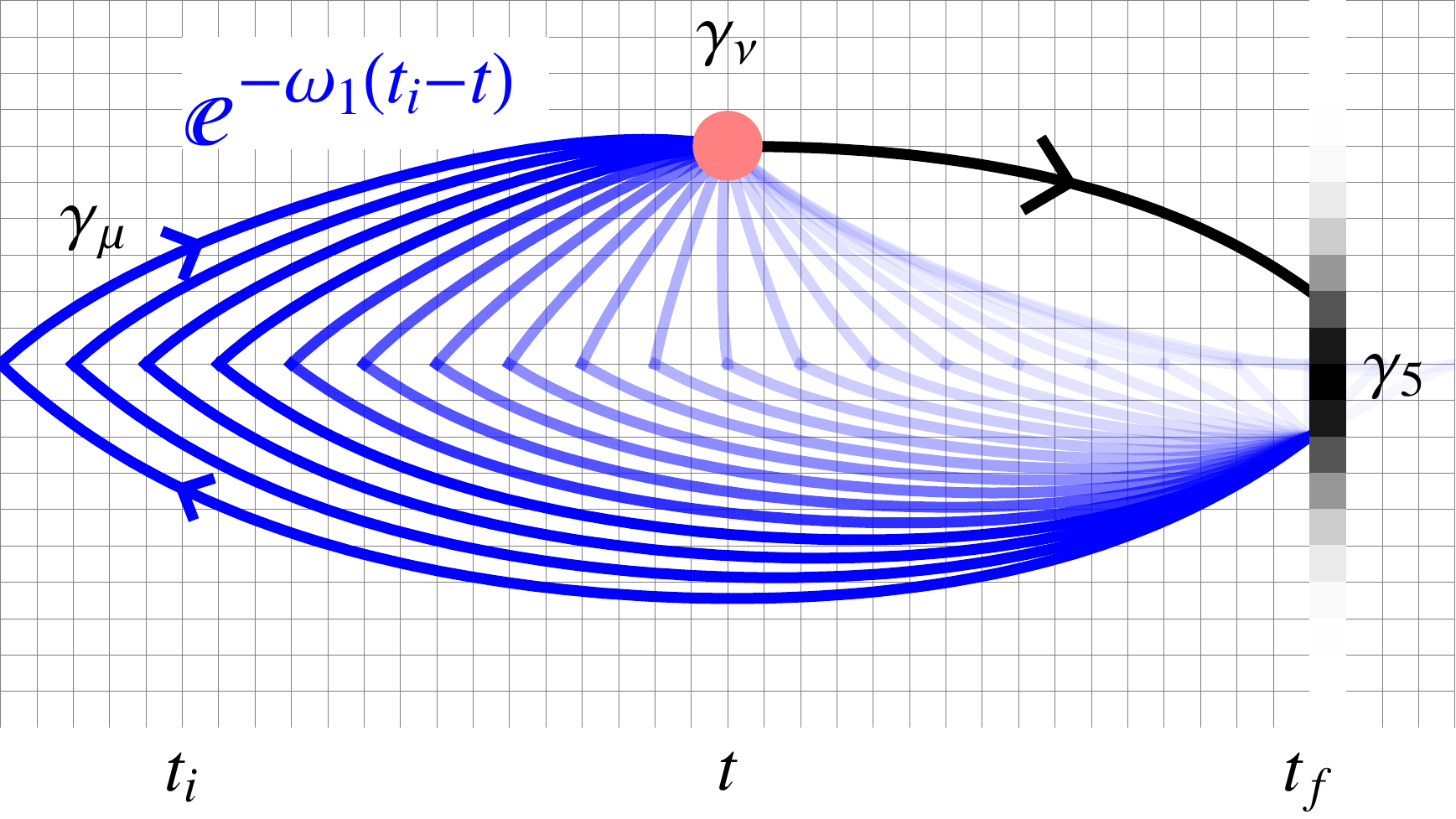}
\caption{{\bf Left:} Schematic of the ``slow'' method. The black arrow indicates an ordinary propagator; the blue arrow is a sequential-source propagator. The three-point function must be evaluated for all $t_i$ and $t$. ${\cal P}$ denotes momentum projection.
{\bf Right:}~Schematic of the ``fast'' method. The sequential propagator source has support on all timeslices and is weighted by the exponential factor. Note that the propagator source is now at $t_f$, reversed relative to the slow method.}
\label{fig:methods}
\end{figure}

For the explicit-integral (``slow'') method, we fix the location of the neutral meson at timeslice $t_f=32$ and zero momentum. We generate propagators using the same clover action used in the dynamical sector under Dirichlet boundary conditions and apply Gaussian smearing to improve overlap with the ground-state mesons. Propagators are generated from point sources at $t_i$; from these, we create sequential-source propagators applying the smearing and momentum-projection at $t_f$. These propagators are then combined into three-point functions by tying them together with various momentum projections ($0\leq|\vec{q_2}|^2\leq5$) at timeslice $t$ and including the appropriate gamma structure for vector currents. (See Figure.~\ref{fig:methods}.)

\begin{figure}[h!]
\includegraphics[width=0.475\textwidth]{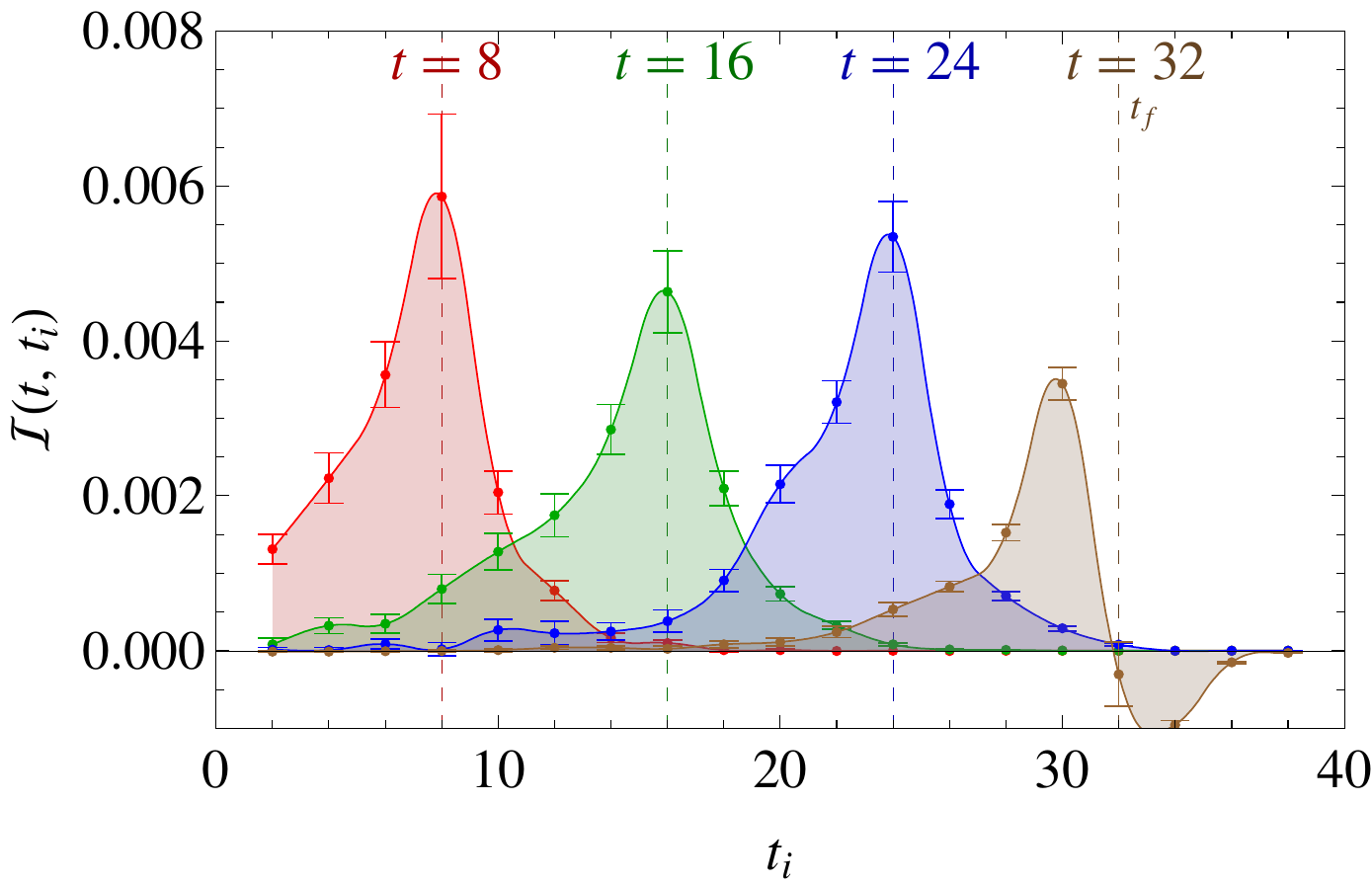}
\hspace{0.05\textwidth}
\includegraphics[width=0.475\textwidth]{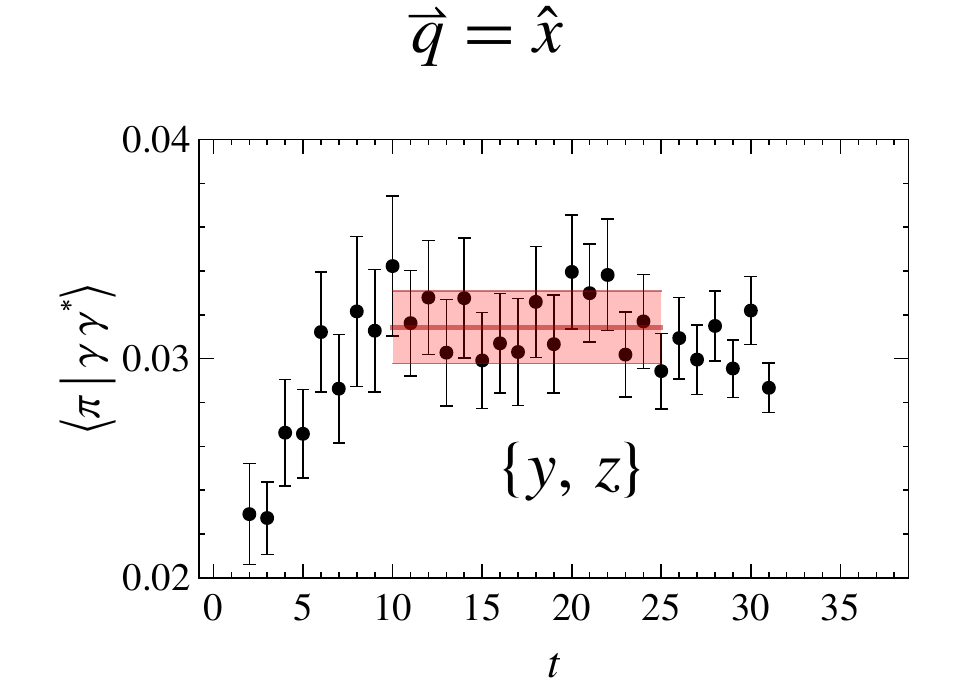}
\caption{{\bf Left:} Dependence of the integrand on selected $t$, denoted by color, and $t_i$, along the x-axis. {\bf Right:}~Integrated matrix element for $\vec{p}\propto\hat{x}$, $\{\mu,\nu\}=\{y,z\}$. Fitted value over the plateau region shown as a pink bar}
\label{fig:time-dependence}
\end{figure}

In order to ascertain whether our calculation will suffer from lattice distortions, we need to scrutinize the time-dependence of the integrand in Eq.~\ref{eq:3pt},
\begin{equation}
\frac{Z_\Phi(p)}{2E_\Phi(p)}\frac{e^{-\omega_1(t_i-t)}}{e^{-E_\Phi(p)(t_f-t)}} {\cal C}_{PVV}(t_f,t,t_i),
\end{equation}
where ${\cal C}_{PVV}$ is the three-point correlator. If the integrand is peaked too sharply, we will not be able integrate it accurately; if it is too wide, we cannot capture the integral within the lattice time extent. In addition, we need to check for distortion due to the boundaries and proximity to the sink timeslice. A sample of our results is shown on the left-hand side of Figure~\ref{fig:time-dependence}. The peak is well resolved, neither too narrow nor too wide; it becomes cut off by the $t_i=0$ boundary for the red $t=8$ peak and is seriously distorted by the $t_f=32$ sink for the brown $t=32$ peak. We must avoid these extremes when looking for a plateau in the integrated matrix element.

We perform the integral explicitly by summing over all $t_i$ for each value of $t$. The integral will be nonzero only when $\varepsilon_{\mu\nu\rho\sigma} \epsilon^\mu \epsilon^\nu q_1^\rho q_2^\sigma \neq 0$. A sample plateau is shown on the right-hand side of Figure~\ref{fig:time-dependence}. As previously discussed, we expect to see a plateau away from the boundary at $t=0$ and the sink at $t=t_f$; we choose to fit to the range $t\in[10,25]$. Although it is possible for there to be exponential contamination from excited states, no such problem is seen here. This may be due to the large gap between the ground-state pion and its first excited state or due to the excited pion having small coupling to the two-photon state.

\begin{figure}[h!]
\includegraphics[width=0.5\textwidth]{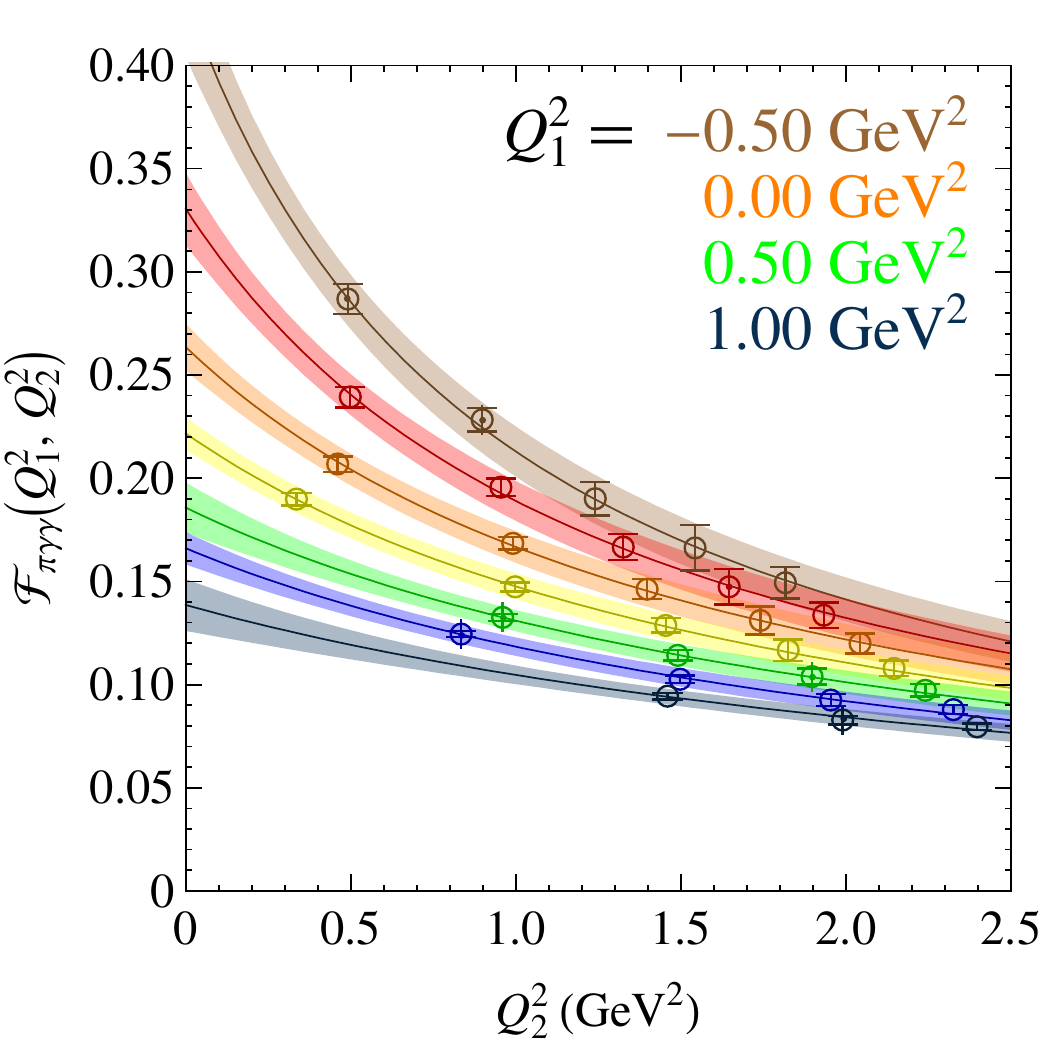}
\includegraphics[width=0.5\textwidth]{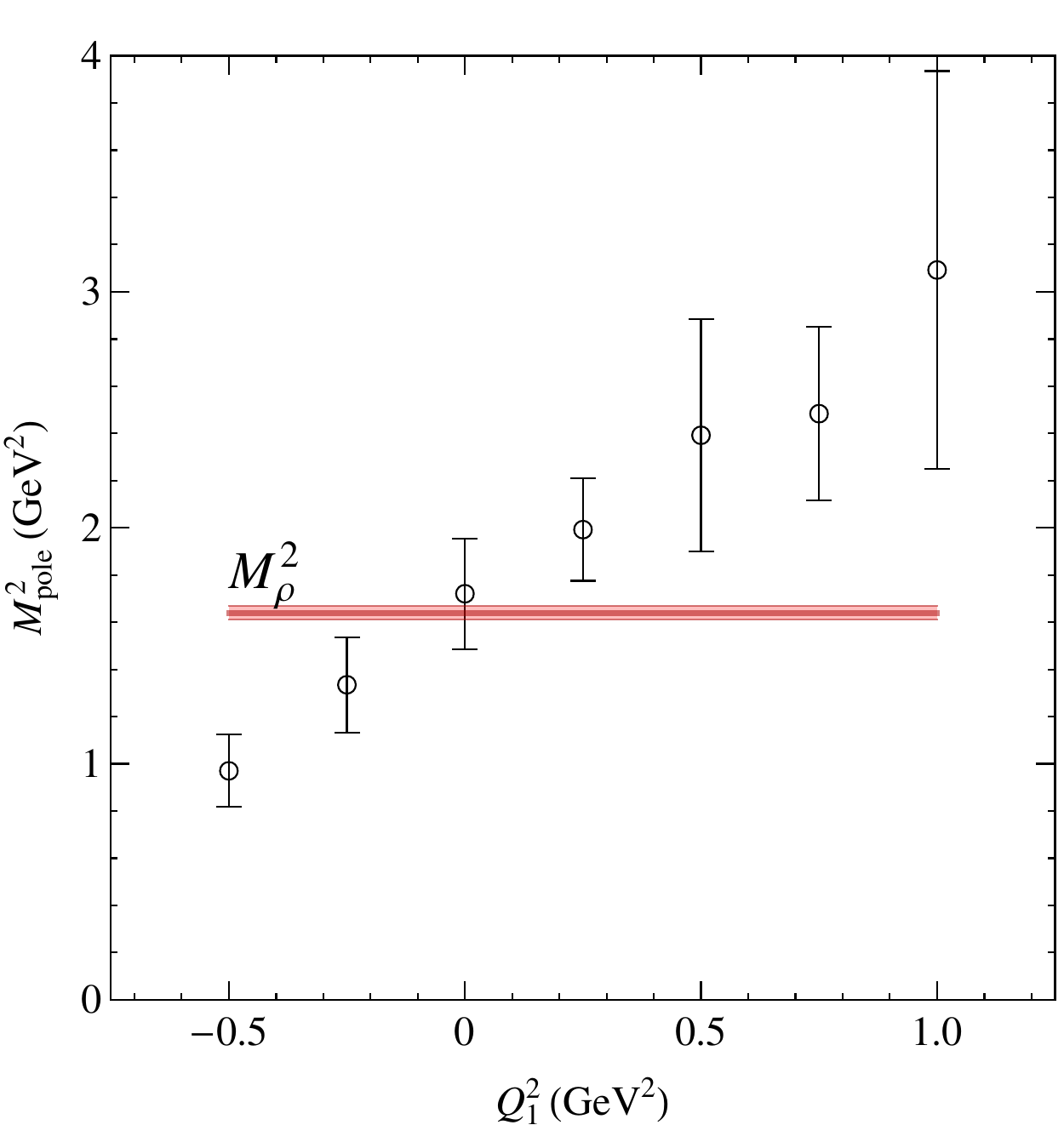}
\caption{{\bf Left:} The neutral meson two-photon decay matrix element as a function of $Q_1^2$ and $Q_2^2$. Different values of $Q_1^2$ (-0.5~GeV$^2$ to 1.0~GeV$^2$ in steps of 0.25~GeV$^2$) are displayed as different colors. $Q_2^2$ runs along the x-axis. The bands show the fit according to the form in Eq.~\protect\ref{eq:monopole}. {\bf Right:}~Pole mass as a function of $Q_1^2$, extracted from fit}
\label{fig:monopole}
\end{figure}

We calculate the neutral pion decay for a variety of photon virtualities. Due to the slow-method construction we use, the value of $\omega_1$ (or equivalently $-Q_1^2 = \omega_1^2 - |\vec{q_1}|^2$) may be set arbitrarily. The value of $Q_2^2$ is then determined by conservation of momentum and energy. We plot the decay constant as a function of $Q_1^2$ and $Q_2^2$ on the left-hand side of Figure~\ref{fig:monopole}. The data are well described by a monopole fit:
\begin{equation}
	{\cal F}(Q_1^2,Q_2^2) = \frac{F(Q_1^2)}{1+Q_2^2/M_{\rm pole}^2(Q_1^2)},
	\label{eq:monopole}
\end{equation}
where $F$ and $M_{\rm pole}$ are fit independently for each $Q_1^2$. Vector-meson dominance (VMD) suggests that the pole mass should be approximately equal to the rho mass. We plot $M_{\rm pole}^2$ as a function of $Q_1^2$ on the right-hand side of Figure~\ref{fig:monopole}. The agreement with the lattice rho-meson mass is relatively good.

\begin{figure}[h!]
\includegraphics[width=0.5\textwidth]{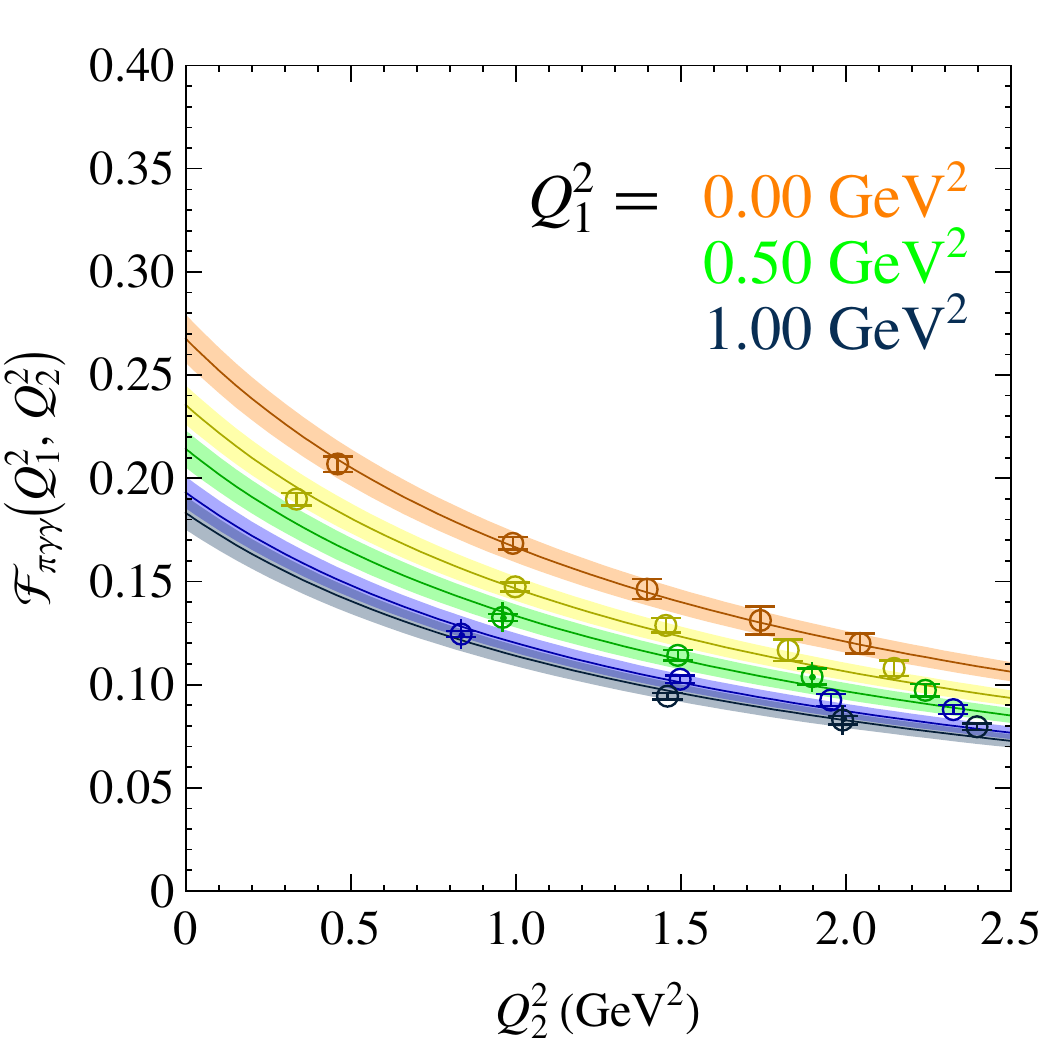}
\includegraphics[width=0.5\textwidth]{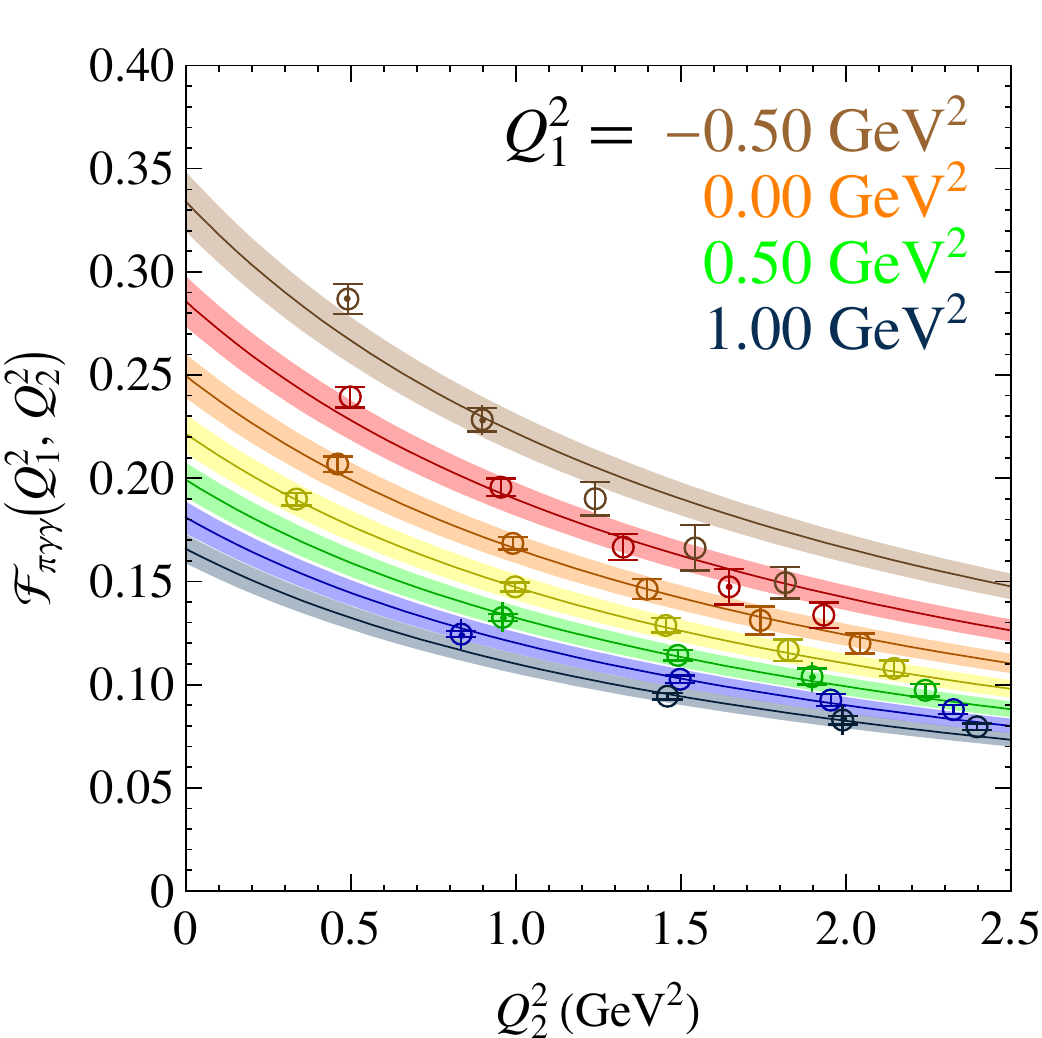}
\caption{As Figure~\protect\ref{fig:monopole} with fits according to {\bf Left:} Eq.~\protect\ref{eq:rhopole} and {\bf Right:}~Eq.~\protect\ref{eq:doublepole}. Only data included in the fit are shown.}
\label{fig:otherpoles}
\end{figure}

If we assume VMD and set the pole mass to $M_\rho$, we can use a simpler form:
\begin{equation}
{\cal F}(Q_1^2,Q_2^2) = \frac{F(Q_1^2)}{1+Q_2^2/M_\rho^2},
	\label{eq:rhopole}
\end{equation}
where $F$ is allowed to vary with $Q_1^2$. We restrict the fit to $Q^2$ regions well away from the pole to avoid distortion. This fit is shown on the left-hand side of Figure~\ref{fig:otherpoles}.

Since the two photons are indistinguishable, there should be symmetry under exchange of $Q_1$ and $Q_2$. We therefore try fitting all the data simultaneously to a double-pole form:
\begin{equation}
	{\cal F}(Q_1^2,Q_2^2) = \frac{F}{\left(1+Q_1^2/M_{\rm pole}^2\right)\left(1+Q_2^2/M_{\rm pole}^2\right)},
	\label{eq:doublepole}
\end{equation}
where now there is only one $F$ and $M_{\rm pole}$. This fit is shown on the right-hand side of Figure~\ref{fig:otherpoles}.
%We note that the goodness of the fit worsens for increasingly negative $Q^2$, probably due to distortion from the on-shell vector meson state at $Q^2=-M_\rho^2$.

This small data sample shows that our method for extracting the matrix element of the neutral meson decay to two photons is working. However, the technique is very expensive, since we must calculate $T$ (the time-extent of the lattice) different propagators to sample the integrand. We can greatly improve this by using a sequential source trick, shown on the right-hand side of Figure~\ref{fig:methods}. Rather than calculating propagators with different $t_i$ individually, we take the solution to the first propagator, multiply it by the exponential factor $e^{-\omega_1(t_i-t)}$ and use this whole quantity as a sequential source. The resulting sequential propagator implicitly contains the integration.

This ``fast'' method has a few disadvantages: We can no longer directly examine the integrand to be sure that we are correctly evaluating the integral. Since we have already invested in a test set using the slow method, this is not a problem. We can also no longer vary $Q_1^2$ without recalculating the sequential propagator, since $\omega_1$ appears in its source. However, since we save a factor of $T$, which can be very large, the fast method will be our method of choice in future work.

%%%%%%%%%%%%%%%%%%%%%%%%%%%%%%%%%%%%%%%%%%%%%%%%%%%%%%%%%%%%%%%%%%%%%%%%%%%%%%%%
\section{Summary}
%%%%%%%%%%%%%%%%%%%%%%%%%%%%%%%%%%%%%%%%%%%%%%%%%%%%%%%%%%%%%%%%%%%%%%%%%%%%%%%%
We have demonstrated that the method of Jung and Ji allows us to access the two-photon decays of neutral mesons in lattice QCD. Using this method on two-flavor lattices, we confirm the predictions of vector-meson dominance, that the matrix element has a monopole form with pole mass approximately equal to the rho mass. In the future, we will apply the fast method to greatly expand the range of quantities we calculate, including conserved-current insertions and scalar and axial meson decays.

%%%%%%%%%%%%%%%%%%%%%%%%%%%%%%%%%%%%%%%%%%%%%%%%%%%%%%%%%%%%%%%%%%%%%%%%%%%%%%%%
\section*{Acknowledgements}
%%%%%%%%%%%%%%%%%%%%%%%%%%%%%%%%%%%%%%%%%%%%%%%%%%%%%%%%%%%%%%%%%%%%%%%%%%%%%%%%
This work was done using the Chroma software suite\cite{Edwards:2004sx} and calculations were performed on clusters at Jefferson Laboratory using time awarded under the SciDAC Initiative. Authored by Jefferson Science Associates, LLC under U.S. DOE Contract No. DE-AC05-06OR23177. The U.S. Government retains a non-exclusive, paid-up, irrevocable, world-wide license to publish or reproduce this manuscript for U.S. Government purposes.

%%%%%%%%%%%%%%%%%%%%%%%%%%%%%%%%%%%%%%%%%%%%%%%%%%%%%%%%%%%%%%%%%%%%%%%%%%%%%%%%
\bibliographystyle{apsrev}
\bibliography{2008LatCon-proc}

\end{document}